\documentclass[prl,showpacs,floats,twocolumn]{revtex4}

\usepackage{graphicx}
\usepackage{amssymb}
\usepackage{amsmath}
\usepackage{hyperref}

\begin{document}

\title{Asymmetric frequency conversion in nonlinear systems driven by a biharmonic pump}
\author{Archana Kamal,$^{1,2 *}$ Ananda Roy,$^{1}$ John Clarke$^{3}$ and Michel H. Devoret$^{1}$}
\affiliation{
$^{1}$Departments of Physics and Applied Physics, Yale University, New Haven, CT 06520, USA\\
$^{2}$Research Laboratory of Electronics, Massachusetts Institute of Technology, Cambridge, MA 02139, USA\\
$^{3}$Department of Physics, University of California, Berkeley, CA 94720-7300, USA}
\email{akamal@mit.edu}
\date{\today}
\begin{abstract}
A novel mechanism of asymmetric frequency conversion is investigated in nonlinear dispersive devices driven parametrically with a biharmonic pump. When the relative phase between the first and second harmonics combined in a two-tone pump is appropriately tuned, nonreciprocal frequency conversion, either upward or downward, can occur. Full directionality and efficiency of the conversion process is possible, provided that  the distribution of pump power over the harmonics is set correctly. While this asymmetric conversion effect is  generic, we describe its practical realization in a model system consisting of a current-biased, resistively-shunted Josephson junction. Here, the multiharmonic Josephson oscillations, generated internally from the static current bias, provide the pump drive.
\end{abstract}
\pacs{85.25.Cp, 85.25.-j, 42.65.Ky}
%85.25.Cp - Josephson devices
%42.65.Ky - Frequency Conversion (nonlinear optics)
%05.45.-a - Nonlinear dynamics
%42.65.Yj - Parametric oscillators and amplifiers, optical
%84.40.Dc - Electronic circuits, microwave
%85.25.-j - Superconducting devices
%74.50.+r - Josephson effect - tunneling phenomena (superconductivity)
\maketitle
Directed transport in nonlinear systems driven by a signal that breaks time-reversal symmetry has received much attention in the last decade \cite{Flach,Fistul2003,PhysRevLett.93.087001,Hanggi2,PhysRevB.84.054515}. This class of systems is closely connected to that of driven systems in which a degree of freedom moves in a potential lacking space inversion symmetry \cite{coldatom}. These various subjects address what is popularly known as the ``ratchet effect"  \cite{Hanggi1}. Here, we focus our attention on a particular type of transport, namely frequency conversion of a signal with the carrier being translated in frequency space either upward (up-conversion process) or downward (down-conversion process). While familiar devices with a purely dispersive nonlinearity, such as 3-wave or 4-wave mixers, can readily perform frequency conversion, reciprocity is maintained between up- and down-converted photon amplitudes with the two directions being distinguished solely by a relative phase determined by the pump tone \cite{ArchanaJPC}.
\par
In this letter, we demonstrate that when the drive contains both the first and second harmonics, either up- or down-conversion can be selected provided the relative phase and amplitude of the tones are set appropriately. Moreover, since this asymmetric conversion process can operate without dissipation in a dispersive nonlinear element such as a superconducting Josephson junction, it can be extended to the quantum regime where signals consist of coherently superposed photon states. Furthermore, a detailed understanding of the conditions under which asymmetric conversion can take place will be useful in the design of quantum information processing  protocols in mesoscopic devices. As an example, we show that asymmetric frequency conversion takes place in the current-biased, resistively shunted Josephson junction (RSJ).
%We argue it is very likely that this particular effect explains why the dc SQUID can be operated as a directional, quantum-limited amplifier.
\par
\begin{figure}[t!]
  % Requires \usepackage{graphicx}
  \includegraphics[width=\columnwidth]{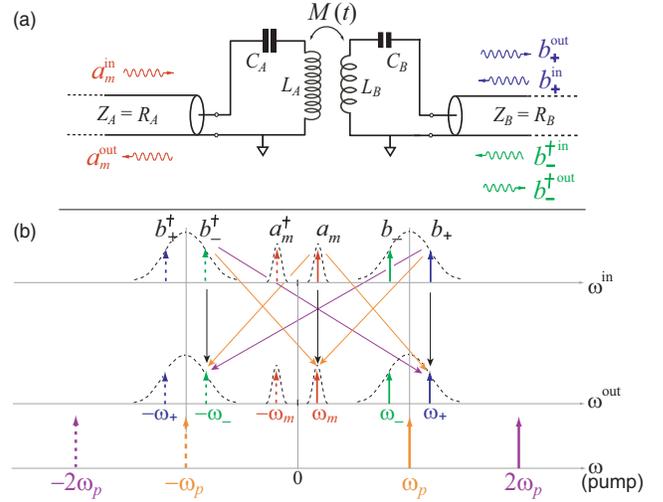}\\
  \caption{Asymmetric parametric frequency converter. (a) Minimal model based on a time-varying mutual inductance $M(t)$ with frequencies $\omega_{p}$ and $2 \omega_{p}$. (b) Frequency landscape showing various frequency mixing processes between the modulation frequency $\omega_{m}$ and the sidebands at $\omega_{\pm}$; solid and dashed arrows indicate  the amplitudes of different modes and their conjugates respectively. The modulation frequency $\omega_{m}$ is chosen to coincide with the center resonance $\omega_{A}$ of the low frequency oscillator, and the pump tone $\omega_{p}$ is chosen to coincide with the center resonance $\omega_{B}$ of the high frequency oscillator. The colors of the mixing arrows indicate the colors of the relevant pump frequencies mediating the process, as shown on the separate pump axis. The black arrows indicate reflections.}
\label{FigParamp}
\end{figure}
To begin with a general discussion, we consider a generic
nonlinear system consisting of two parametrically coupled
oscillators $A$ and $B$. We model the parametric coupling as a
time-varying mutual inductance $M(t)$, as shown in Fig.
\ref{FigParamp}(a). Varying this coupling at the pump frequency
$\omega_{p}$ at the difference or sum of the two coupled modes
results in three-wave mixing, leading to either frequency
conversion or amplification respectively \cite{Louisell}. In
microwave circuits such a coupling scheme may be implemented by
employing the nonlinear Josephson inductance, $L_{J}(I) = L_{J}(0)
[1- ( I/I_{0})^{2}]^{-1/2}$. When the current $I(t) = i_{rf} \cos
(\omega_{p} t)$ flowing through the junction is much smaller than
the critical current $I_{0}$, the junction behaves as an effective
time-varying inductance $L_{J}(t) \approx L_{J}(0) I(t)/I_{0}$
modulated at the pump frequency by the rf current $I(t)$. In Fig.
\ref{FigParamp}(b), we show a more generic configuration
\cite{circNature} where the pump frequency $\omega_{p}$ is aligned
with the oscillator resonance $\omega_{B}$, fulfilling the
conversion condition ($\omega_{P} = \omega_{+} - \omega_{A}$) with
respect to the upper sideband at $\omega_{+}$ and the
amplification condition ($\omega_{P} = \omega_{-} + \omega_{A}$)
with respect to the lower sideband at $\omega_{-}$.
\par
A convenient formalism to capture the dynamics of such parametric systems is provided by input-output theory \cite{IOT}. In this paradigm, a three-wave mixer is fully described by a scattering matrix that gives the relationship between each incoming and outgoing mode of the system,
\begin{eqnarray}
    \left(\begin{array}{c}
    a^{\rm out} [\omega_{m}]\\
    b^{\rm out} [\omega_{+}]\\
    b^{\rm out} [-\omega_{-}]
    \end{array} \right)
    =
    \left(\begin{array}{ccc}
    r_{m} & t_{d} & s_{d} \\
    t_{u} & r_{+} & v_{+-}\\
    s_{u} & v_{-+} & r_{-}
    \end{array} \right)
    \left(\begin{array}{c}
    a^{\rm in} [\omega_{m}]\\
    b^{\rm in} [\omega_{+}]\\
    b^{\rm in} [-\omega_{-}]
    \end{array} \right),\;\;
\end{eqnarray}
where different frequency components participating in the mixing
process, $\omega_{m}$ and $\omega_{\pm}=  \omega_{p} \pm
\omega_{m}$, are selected by harmonic balance. The $a$ and $b$
denote incoming and outgoing wave amplitudes for spatial channels
$A$ and $B$, normalized in terms of single photon energies, such
that $\langle a^{\rm in, out} [\omega] a^{\rm in, out}[\omega']
\rangle = S_{a^{\rm in, out}a^{\rm in, out}}
[(\omega-\omega')/2]\delta (\omega + \omega')$ where $S_{a^{\rm
in, out}a^{\rm in, out}}[\omega]$ represents the number of photons
per unit time per unit bandwidth around frequency $\omega$.
Negative frequency amplitudes denote conjugation along with
conversion; for real signals $a[-\omega] = a^{\dagger} [\omega]$.
Here $t_{u}, \; s_{u}$ describe up-conversion from the modulation
frequency $\omega_{m}$ to sideband frequencies $\pm\omega_{p} +
\omega_{m}$ while $t_{d}, \; s_{d}$ describe the reverse process
of down-conversion from sideband frequencies to the modulation
frequency. The diagonal elements denoted by $r_{m, +, -}$ are the
respective reflection coefficients, and the $v_{\pm\mp}$
coefficients denote the strength of mixing within the sidebands.
The frequency landscape in Fig. \ref{FigParamp}(b) depicts these
processes pictorially. All the three-wave mixing processes
considered here are assumed to be phase matched since the mixing
process is spatially local in the lumped element geometry shown in
Fig. \ref{FigParamp}. Hence, the phases of the signals are entirely
determined by the phase of the corresponding pump mediating the
mixing process. In addition, the oscillation amplitudes at signal and
sideband frequencies are considered to be weak enough to be in the
linear input-output regime, where instabilities and nonlinearities
due to pump depletion \cite{ArchanaJPC, Kaup} can be safely
ignored.
\par
We now consider two distinct pumping schemes.
\\
\emph{Case I}: $M (t) = M_{1} \cos(\omega_{p} t)$.
This corresponds to the usual case of parametric pumping with a monochromatic tone. Following an analysis similar to that in \cite{circNature}, we obtain
\begin{subequations}
\begin{align}
    &t_{d} = t_{u} = 2 i \xi_{1}, \\
    & s_{d} = s_{u}^{*} = 2 i \xi_{1},\\
    &  v_{+-} = 2 \xi_{1}^{2}; \quad v_{-+} = -2 \xi_{1}^{2},
\end{align}
\label{EqSymFC}%
\end{subequations}
where $\xi_{1} = (M_{1}/\sqrt{L_{A} L_{B}})$ is the equivalent
dimensionless pump strength \cite{footnote1}. Thus, we recover the
usual case of symmetric (or reciprocal) frequency conversion with
scattering between any pair of signals of equal magnitude, i.e.
$|t_{d}| =|t_{u}|, \;|s_{d}| = |s_{u}|$. We note, however, that
the phases associated with the off-diagonal scattering elements
are not equal. For mathematical simplicity, we work under the
rotating wave approximation (RWA) and assume that both oscillators
$A$ and $B$ are driven near resonance, i.e. $\omega_{m} =
\omega_{A}$ and $\delta_{\pm} = (\omega_{\pm} -
\omega_{B})/\Gamma_{B} \ll1$ ($\Gamma_{B} = R_{B}/ (2 L_{B})$ is
the linewidth of the high frequency oscillator).
\\
\emph{Case II}: $M(t) = M_{1} \cos(\omega_{p} t) + M_{2} \cos(2\omega_{p} t + \alpha)$.
In the case of biharmonic driving, additional three-wave mixing of sidebands by the $2 \omega_{p}$ harmonic modifies the scattering amplitudes to (assuming $\delta_{\pm} =0$)
\begin{subequations}
\begin{align}
    & t_{d} = 2 i \xi_{1} D (1 + i  e^{-i \alpha} \xi_{2}); \quad
    t_{u}  = 2 i \xi_{1} D (1 + i  e^{i \alpha} \xi_{2}),\\
    & s_{d} =2 i \xi_{1}  D(1 - i  e^{i \alpha} \xi_{2});  \quad
    s_{u}  = - 2 i \xi_{1}  D(1 - i e^{-i \alpha}  \xi_{2}),\\
   &v_{+-} = 2  D (\xi_{1}^{2} +  i e^{i \alpha}  \xi_{2}); \quad
    v_{-+} =  -2 D (\xi_{1}^{2} +  i e^{-i \alpha}  \xi_{2}),
\end{align}
\label{EqNRfreqconv}%
\end{subequations}
\begin{figure}[t!]
  % Requires \usepackage{graphicx}
  \includegraphics[width=\columnwidth]{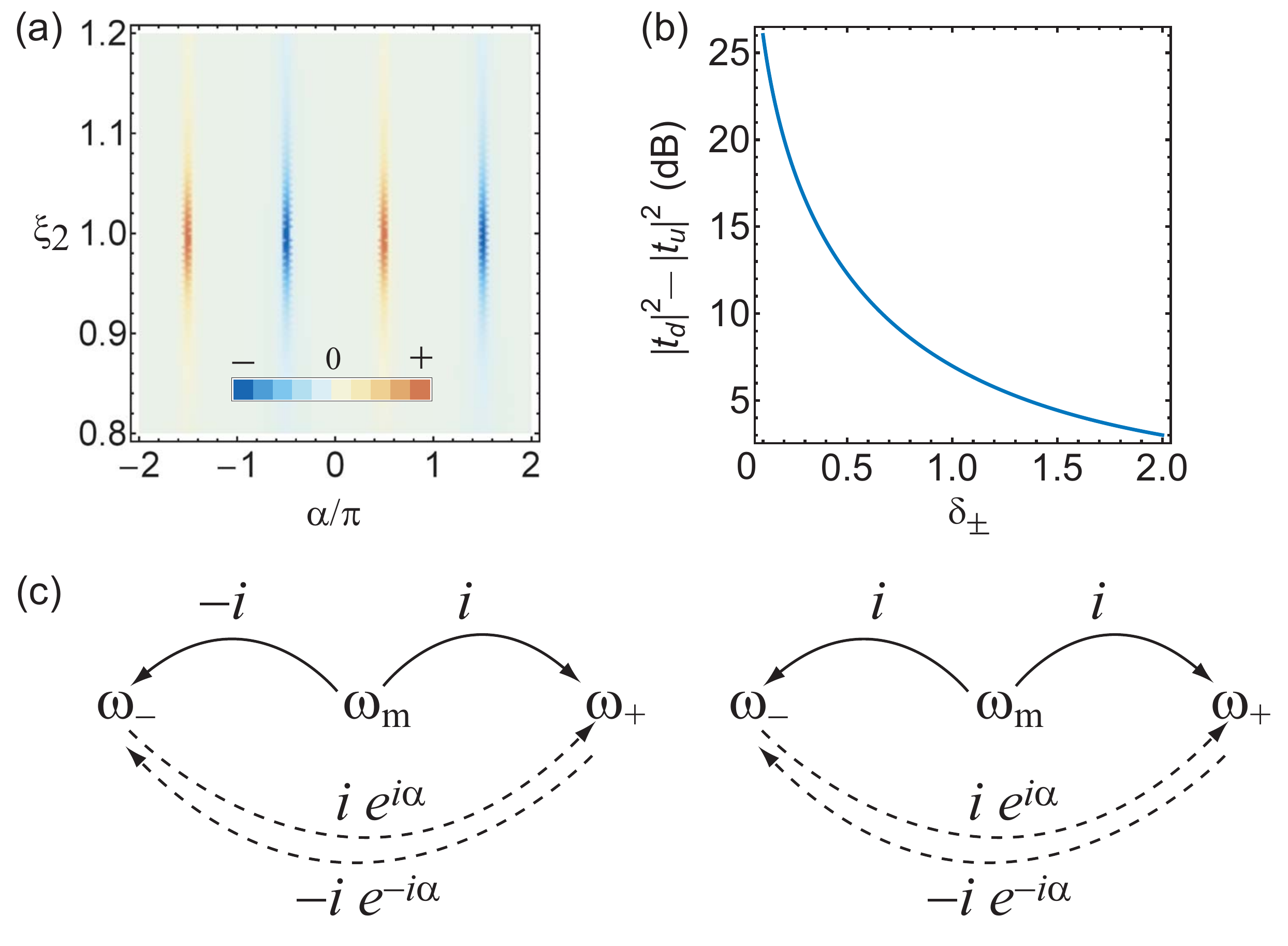}\\
  \caption{Asymmetric frequency conversion. (a) Asymmetry between frequency conversion amplitudes $|t_{d}|^{2} -|t_{u}|^{2}$ [Eq. (\ref{EqNRfreqconv})] plotted as a function of the strength of the second harmonic ($\xi_{2}$) of the pump and phase frustration $\alpha$ between the harmonics. Maximum asymmetry is realized for maximal frustration $\alpha = \pi/2 \pm n\pi$, with down-conversion occurring when the second harmonic leads the first harmonic ($\alpha = + \pi/2 + 2 n \pi$) and up-conversion occurring when the second harmonic lags behind the first harmonic  ($\alpha = - \pi/2 + 2 n \pi$). (b) Maximal asymmetry plotted as a function of modulation frequency parametrized as dimensionless detuning $\delta_{\pm} = \omega_{m}/\Gamma_{B}$. (c) Interference of different scattering amplitudes showing the realization of asymmetry as in (a). Solid lines indicate the usual frequency conversion in the presence of a single monochromatic pump at frequency $\omega_{p}$ while the dashed arrows indicate additional processes mediated by its second harmonic $2 \omega_{p}$.}
\label{FigAsym}%
\end{figure}
where $\xi_{2} =  (M_{2}/\sqrt{L_{A} L_{B}})$ and $D=[2 i \xi_{1}^{2} \xi_{2} \cos\alpha$ ${+ (1- \xi_{2}^{2})]^{-1}}$. Equations (\ref{EqNRfreqconv}) show that the usual symmetry of frequency conversion is now broken, since $|t_{d}| \neq |t_{u}|$ and $|s_{d}| \neq |s_{u}|$. Moreover, the preferred conversion channel is guided by $\alpha$, the phase shift between the harmonic components of the drive. Maximum asymmetry between off-diagonal scattering coefficients is obtained for $\alpha = \pm \pi/2$, with $+\pi/2$ yielding down-conversion and $-\pi/2$ yielding up-conversion (Fig. \ref{FigAsym}). For zero phase difference between the two harmonics, $\alpha = 0\; ({\rm mod}\; 2 n \pi)$, we recover symmetric scattering, although with modified coefficients because of additional 3-wave mixing by the $2 \omega_{p}$ pump. The physical mechanism underlying this asymmetry is rooted in the interference of different conversion paths in frequency space between a given pair of modes. As shown in Fig. \ref{FigAsym}(c), the phases of different scattering coefficients are governed by the phases of respective harmonics mediating the process. The introduction of three-wave mixing processes between the sidebands $\omega_{\pm}$ closes the interference loop; the phase sensitive nature of up- and down-conversion coefficients ($t, \,s$), obtained with the single-frequency pump  in \emph{case I}, is translated through these processes into an asymmetry between off-diagonal scattering amplitudes.
\par
Biharmonically pumped parametric systems form an intriguing
platform for exploring qualitatively new physics, the asymmetric
frequency conversion discussed here being one such example. Steady
state properties near the critical bifurcation threshold in such
systems have also been studied in detail previously
\cite{Drummondone}, with photon statistics displaying
bunching/antibunching depending on the relative phase between the
two pump drives \cite{Drummondtwo}. Furthermore, in conjunction
with the recently proposed ideas of reservoir engineering
\cite{MetelmannPRL}, such parametric conversion protocols can be
shown to implement quantum-limited directional amplification
\cite{MetelmannKamal, Ranzani}.
\begin{figure}[t!]
  % Requires \usepackage{graphicx}
  \includegraphics[width=\columnwidth]{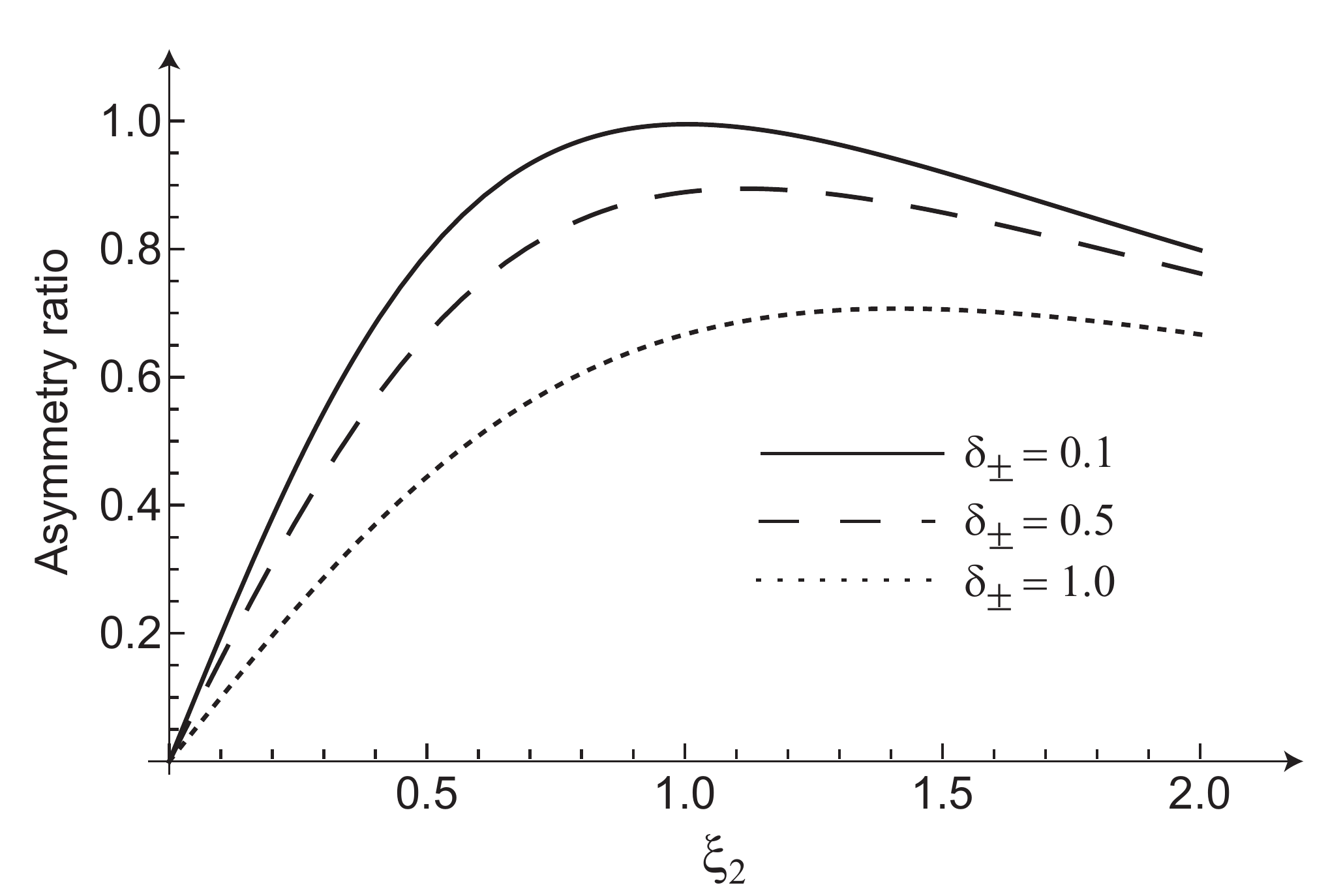}\\
  \caption{Asymmetry ratio defined as the net difference in up-converted  and down-converted power, normalized to the total converted power $(|t_{d}|^{2}+|s_{d}|^{2} -|t_{u}|^{2}-|s_{u}|^{2})/(|t_{d}|^{2}+|s_{d}|^{2} +|t_{u}|^{2}+|s_{u}|^{2})$, plotted as a function of the strength of the second harmonic $\xi_{2}$, for three different detunings (or modulation frequencies). This ratio is independent of the pump strength $\xi_{1}$. The maximum ratio of unity, corresponding to 100\% asymmetry, is achievable for optimal strengths of $2 \omega_{p}$ pump and sufficiently small detunings from the carrier ($\omega_{m} \ll \omega_{p}$).}
\label{FigAsymRatio}%
\end{figure}
\par
As an example of a multiharmonic-pump mediated frequency conversion, we now consider in detail the scattering of a microwave signal by an RSJ current biased in its voltage state. The RSJ has been studied extensively both theoretically \cite{PhysRevB.50.395,SQUIDvol1} and experimentally \cite{RKoch,JoyezJTJ}. The phenomenon of directed transport in a single-junction functioning as fluxon ratchets has also received attention \cite{PhysRevLett.95.090603,PhysRevLett.104.190602}, though these studies have focussed on the static current-voltage characteristics. On the other hand, studies of mixer properties of the RSJ have remained restricted to considering Josephson oscillations as the carrier or local oscillator \cite{Likharev}. Detailed studies of higher Josephson harmonics in the RSJ have remained elusive, to the best of our knowledge, and the emergence of dynamical nonreciprocity under their influence has remained hitherto unexplored. Here, we present a generalized input-output formalism that enables a self-consistent analytical evaluation of amplitudes and phases of Josephson harmonics, generated internally by the ac Josephson effect, to any order.
\par
We begin with the well known nonlinear equation of motion for a Josephson junction biased with a static current $I_{B}$,
\begin{eqnarray}
    \frac{\varphi_{0}}{R} \dot{\varphi} + I_{0}\frac{\partial}{\partial \varphi}\left(\frac{U}{E_{J}}\right) = 0,
    \label{EqRSJ}
\end{eqnarray}
representing a particle moving in the washboard potential $U(\varphi)= E_{J} \left(1-\cos\varphi - \varphi I_{B}/I_{0}\right)$. Here, $\varphi$ denotes the phase difference across the junction, $E_{J}= \varphi_{0} I_{0}$ is the Josephson energy, $I_{0}$ the critical current, $\varphi_{0} = \hbar/(2e)$ the reduced flux quantum and $R$ the shunt resistance. The presence of the $\cos\varphi$ term  in $U(\varphi)$ causes the dynamics to be highly nonlinear, generating multiple Josephson harmonics for $I_{B}> I_{0}$. These harmonics play a role analogous to that of the multiharmonic pump discussed earlier. To describe the mixing properties of the RSJ, we perform a perturbative analysis by expressing the phase as $\varphi = \omega_{J} t + \delta\varphi (t)$ and expanding the $\cos\varphi$ term in a series in $\delta\varphi$ about $\omega_{J} t$, where $\omega_{J} = \langle\dot{\varphi}\rangle$ is the Josephson frequency.
\par
\begin{figure}[t!]
  % Requires \usepackage{graphicx}
  \includegraphics[width=\columnwidth]{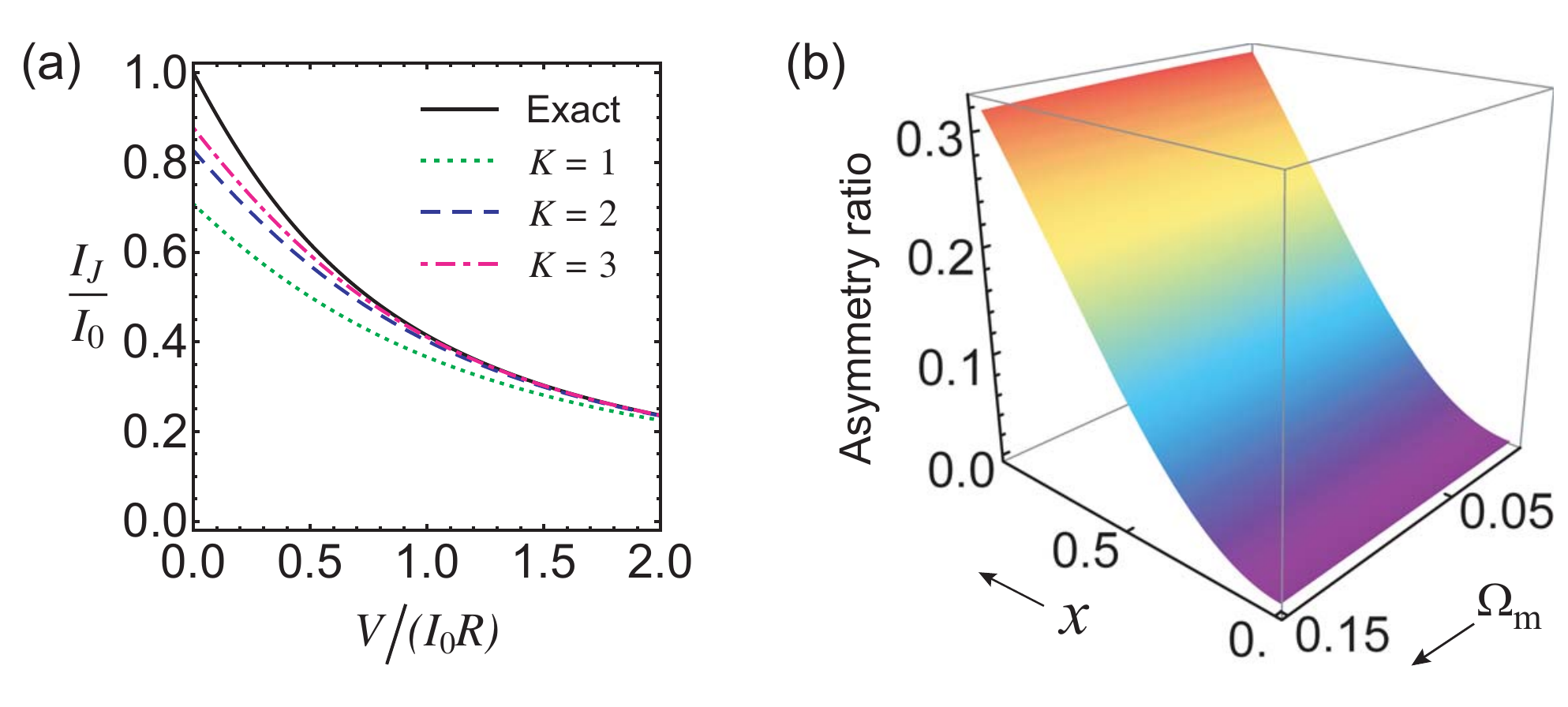}\\
  \caption{Analytical calculation for the RSJ. (a)  Static $I-V$ characteristics of the RSJ calculated using the perturbative series method, showing the variation of voltage $V$ across the junction with current $I_{J}$ flowing through it. The solid black line corresponds to the exact analytical result $V = R \sqrt{I_{B}^{2}- I_{0}^{2}}$, while the green (dotted), blue (dashed) and violet (dotted-dashed) curves correspond to the series calculation with one [$K=1$], two [$K=2$] and three [$K=3$] Josephson harmonics, respectively. (b) Asymmetry ratio (cf. Fig. \ref{FigAsymRatio}) for the RSJ, calculated with the pump configuration derived in Eq. (\ref{Eqpumps}). This ratio is almost independent of the modulation frequency (parameterized here as a dimensionless variable $\Omega_{m} = \omega_{m}/\omega_{B}$ with $\omega_{B} = I_{B} R/\varphi_{0}$) for small $\Omega_{m}$, and increases as the junction is biased towards the strongly nonlinear regime by increasing $x$ (or decreasing $I_{B}$) \cite{Supplement}.}
\label{FigRSJ}
\end{figure}

This treatment allows us to write
\begin{eqnarray}
     \delta\varphi (t) = \sum_{k=1}^{K} p_{k}^{I} \cos(k \omega_{J}t) + p_{k}^{Q} \sin(k \omega_{J}t)
\label{Eqrfcomps}
\end{eqnarray}
in terms of the harmonic components associated with the junction dynamics. Here, $(p_{k}^{I},  p_{k}^{Q})$ denote the amplitudes of the two quadratures associated with the $k^{\rm th}$ harmonic of $\omega_{J}$. Following the analysis scheme introduced earlier \cite{ArchanaSQUID}, we evaluate $p_{k}^{I,Q}$  for different $k$-values as a truncated power series in the dimensionless inverse bias parameter $x =I_{0}/I_{B}$, with the number of Josephson harmonics $K$ included in the calculation set by the order of expansion of $\cos\varphi$. A self-consistent calculation of the static $I-V$ characteristics, which are related to the amplitudes of the Josephson oscillation and its harmonics, is presented in Fig. \ref{FigRSJ}(a).
\par
The calculated amplitude of the Josephson oscillation using this method to first order [corresponding to $K=1$ in Eq. (\ref{Eqrfcomps})] is  \cite{Supplement}
\begin{eqnarray*}
    p_{1}^{I} = x; \quad p_{1}^{Q} =0.
    \label{pumpvals}
\end{eqnarray*}
Similarly, including  the next order in the perturbation series [corresponding to $K=2$ in Eq. (\ref{Eqrfcomps})] yields the following expressions for the amplitudes of the first and second Josephson harmonics in terms of the bias parameter $x$:
\begin{subequations}
\begin{align}
    p_{1}^{I} &= x + \frac{x^{3}}{4}; \quad
    p_{1}^{Q} = 0, \\
    p_{2}^{I}  &= 0; \quad
    p_{2}^{Q} = -\frac{x^{2}}{4}.
\end{align}
\label{Eqpumps}%
\end{subequations}
Extending the analysis to higher orders, we find the entire series of Josephson harmonics generated by a junction in its voltage state to be
\begin{eqnarray}
    \delta\varphi (t) = \sum_{k=1}^{\infty} p_{2k-1}^{I} \cos[(2k-1)\omega_{J}t] + p_{2k}^{Q} \sin[(2k)\omega_{J}t]. \nonumber
\label{EqRSJpumps}
\end{eqnarray}
Such a series, with the phases of alternate even and odd harmonics shifted by $\pi/2$, is frequently encountered in nonlinear systems, for example driven ratchets, leading to directed spatial transport \cite{Hanggi1}. A similar drive configuration has also been explored to study directed diffusion in spatially periodic potentials, implemented by the tuning of activation energies with phase between the two harmonics \cite{DykmanPRL}. In the particular case considered here, it corresponds to the phase across the junction evolving in a \emph{time-asymmetric sawtooth} fashion.
\par
As discussed for a general parametric system in the previous section, we can now study the nonlinear mixing properties of the RSJ under such time-asymmetric phase evolution. This is accomplished by treating the shunt resistor as a semi-infinite transmission line which serves as a channel for incoming and outgoing waves at the modulation frequency $\omega_{m}$ and sideband frequencies $\omega_{\pm} = \pm n \omega_{J}+ \omega_{m}$ of interest \cite{footnote2}. We model the junction response to such input drives by including a perturbative radiofrequency (rf) component of the form $\delta\varphi_{S} (t) = \sum_{n=-N}^{+N} s^{I} \cos[(n\omega_{J}+ \omega_{m})t] + s^{Q} \sin[(n\omega_{J} +\omega_{m})t]$ in Eq. (\ref{Eqrfcomps}) and performing a harmonic balance analysis at $\omega_{m}$ and $\omega_{\pm}$. The combination of the Josephson harmonics evaluated earlier plays the role of an effective multiharmonic pump. Figure \ref{FigRSJ}(b) is a plot of the relative conversion asymmetry, obtained as a difference between the net down- and up-converted amplitudes normalized to the total converted power. This asymmetry ratio is calculated by restricting the perturbative analysis to the second harmonic ($K=2$) and the first pair of sidebands  ($N=1$) generated about the Josephson frequency [in accordance with the frequency landscape shown in Fig. \ref{FigParamp}(b)]. The figure shows that the scattering becomes nonreciprocal, with the coefficients describing down-conversion stronger than those describing up-conversion. This result is in agreement with that obtained for a general parametric mixing scheme in view of the pump configuration described by Eq. (\ref{Eqpumps}). Since $\alpha = +\pi/2$, the second Josephson harmonic leads the first harmonic and, as described previously, such a pump configuration leads to down-conversion (see Fig. \ref{FigAsym}). We note that the maximum asymmetry ratio obtained for the RSJ is around 0.3; this is because this ratio depends crucially on the strength of the $2\omega_{J}$ harmonic (Fig. \ref{FigAsymRatio}), which in the case of RSJ is generated by a higher order process and is consequently weaker than the $\omega_{J}$ pump [Eq. (\ref{Eqpumps})].  These results are corroborated qualitatively by those obtained through a direct numerical integration of Eq. (\ref{EqRSJ}) \cite{Supplement}. It may be possible to employ filters across the junction to modify the junction reactance near the Josephson frequency and its harmonics, and achieve a more favorable proportion of higher Josephson harmonics conducive to higher conversion asymmetry.
\par
In conclusion, we have uncovered a generic mechanism that breaks the symmetry of frequency conversion in parametric processes. Unlike the symmetric conversion schemes employing a single frequency pump, it relies on a multiharmonic pump with appropriate relative phases and amplitudes of successive harmonics that select frequency up-conversion over down-conversion, or vice versa, with 100\% efficiency.  We have shown that the asymmetry in frequency conversion takes place in a current-biased RSJ. The internally generated series of Josephson harmonics has odd and even harmonics phase-shifted by $\pi/2$, resulting in a pump configuration optimally tuned for down-conversion. The asymmetric frequency conversion protocol in the RSJ also provides important clues for unravelling the operation of the dc SQUID (Superconducting QUantum Interference Device) \cite{SQUIDvol1} as a directional, quantum-limited, microwave amplifier \cite{ArchanaSQUID, Ranzani}. The SQUID may be modeled as a two-variable RSJ circuit with a static phase difference between the two junctions imposed by the external flux bias of the loop \cite{ArchanaSQUID}.
\par
Given its generic platform-independent nature, our analysis opens itself to various applications. In addition to being \emph{in situ} tunable with pump phases, the parametrically guided frequency conversion can easily be extended to a multi-photon regime without being limited by saturation effects such as those encountered in down-conversion schemes based on three-level $\Lambda$ systems \cite{Koshino1}. Selective down-conversion can also be used to enhance the generation efficiency of nonclassical states such as entangled photons \cite{SPDC}. Furthermore, the protocol of efficient up-conversion with gain may complement recent proposals \cite{Bagci} for long distance transmission of quantum information using low loss optical technology.
\par
The authors thank J. Aumentado, S. M. Girvin, A. J. Kerman, R. J. Schoelkopf and A. D. Stone for helpful discussions. A. K. would especially like to thank A. Metelmann and L. Ranzani for useful insights and N. Masluk for help with numerical simulations. This research was supported by the NSF under Grant No. DMR-1006060 and the NSA through ARO Grant No. W911NF-09-1-0514.
%
%%%%%%%%%%%%%%%%%%%%%%%%%%%%%%%%%%%%%%%%%%%%%%%%%%
%
%
%\bibliography{RSJbib}

\begin{thebibliography}{34}
\bibitem[{foo({\natexlab{a}})}]{footnote1}
\bibinfo{note}{The mixing between sidebands here is a second order, four-wave
  mixing process, as indicated by the exponent of the drive term entering the
  coefficients $v_{\pm\mp}$.}
\bibitem[{foo({\natexlab{b}})}]{footnote2}
\bibinfo{note}{Unlike the parametric model considered earlier, in the
  transmission line model for the RSJ we can assume that both the signal and
  sideband waves travel on the same spatial port as there is no resonant
  structure in this model. This distinction is not crucial to the problem at
  hand as the form of the scattering coefficients is not affected by this
  change, up to a common complex factor, preserving the relative phases between
  them.}
\end{thebibliography}
%\bibliographystyle{apsrev}
%
%

%
\end{document}

\title{Supplemental Material for\\
``Asymmetric frequency conversion in nonlinear systems driven by a biharmonic pump"}
\author{Archana Kamal,$^{1,2}$ Ananda Roy,$^{1}$ John Clarke$^{3}$ and Michel H. Devoret$^{1}$}
\affiliation{
$^{1}$Departments of Physics and Applied Physics, Yale University, New Haven, CT 06520, USA\\
$^{2}$Research Laboratory of Electronics, Massachusetts Institute of Technology, Cambridge, MA 02139, USA\\
$^{3}$Department of Physics, University of California, Berkeley, CA 94720-7300, USA}

\date{\today}
\maketitle
%
%
%cccccccccccccccccccccccccccccccccccccccccccccccccccccccccccc
\section{Steady state response and Josephson harmonics in the RSJ}
%cccccccccccccccccccccccccccccccccccccccccccccccccccccccccccc
%
In this section, we present a detailed calculation of the steady-state amplitudes of the Josephson oscillation and its harmonics ($\omega_{J}, 2\omega_{J}$ ..)  in an RSJ. This requires a self-consistent evaluation of the static and rf characteristics of the junction, since through the second Josephson relation $V/(I_{0} R) = \omega_{J}/\omega_{0}$ (where $\omega_{0} = I_{0} R/\varphi_{0}$), there exists an intimate connection between the Josephson oscillation frequency and the static voltage that develops across the junction in the running state.
\par
For evaluating the steady state response of the RSJ, we consider Eq. (4) of the main text with
\begin{eqnarray}
    \varphi (t) = \omega_{J} t + \delta\varphi (t),
\end{eqnarray}
where $\delta\varphi (t)$ denotes the oscillatory part of the
evolution due to Josephson oscillations (see Eq. (5) of the main
text).  As explained in the main text, we consider the effect of
these internally generated oscillations on the phase evolution
perturbatively by expanding the junction characteristics as a
truncated series in $\delta\varphi(t)$ about the operating point
set by $\omega_{J}t$. Furthermore, for computational efficiency, we
expand both the Josephson oscillation amplitudes and the
Josephson frequency corresponding to static voltage across the
junction, in terms of the inverse bias parameter $x = I_{0}/I_{B}
= \omega_{0}/\omega_{B}\ll 1$ \footnotemark \;as \footnotetext{The
symbol $\omega_{B} \equiv I_{B} R/\varphi_{0}$ is a
parametrization of the bias current on a frequency scale.}
\begin{eqnarray}
    & & p_{k}^{I, Q} = \sum_{a=0}^{2K -1} p_{k,a}^{I,Q} x^{a}\\
    \label{Eqpumpsexample}
    & & \frac{V}{I_{B}R} = \frac{\omega_{J}}{\omega_{B}} \equiv \sum_{b=0}^{2K} v_{b} x^{b},
\end{eqnarray}
where $K$ denotes the number of Josephson harmonics included in the analysis.
\par
For illustrative purposes, let us consider the $K=1$ (first Josephson harmonic) case. First, we use Eq. (\ref{Eqpumpsexample}) to calculate the voltage term in Eq. (4), and obtain
\begin{eqnarray}
\dot{\varphi}= \omega_{J} [1- ( p_{1,0}^{I}+ p_{1,1}^{I} x) \sin\omega_{J} t + ( p_{1,0}^{Q}+ p_{1,1}^{Q} x) \cos\omega_{J} t]. \nonumber\\
\label{Eqvoltageexample}
\end{eqnarray}
Similarly, for the current through the junction, we obtain
\begin{eqnarray}
    \omega_{0} \sin\varphi &=& \omega_{0} \sin \omega_{J} t + \omega_{0}\cos (\omega_{J}t) [ ( p_{1,0}^{I}+ p_{1,1}^{I} x) \cos\omega_{J} t \nonumber\\
& & \qquad \qquad+ ( p_{1,0}^{Q}+ p_{1,1}^{Q} x) \sin\omega_{J} t].
\nonumber\\
\label{Eqcurrentexample}
\end{eqnarray}
Using Eqs. (\ref{Eqcurrentexample}) and (\ref{Eqvoltageexample}) in Eq. (4) and collecting terms oscillating at $\omega_{J}$, we obtain
\begin{eqnarray}
    (p_{1,0}^{I} + p_{1,1}^{I} x) v_{0} = x \\
    p_{1,0}^{Q} + p_{1,1}^{Q} x = 0,
\end{eqnarray}
where we have separated the contributions of the two quadratures to the amplitude of the harmonic. This gives, $p_{1,0}^{I} = 0, \;p_{1,1}^{I} = v_{0}^{-1}, \;p_{1,0}^{Q} = 0, \; p_{1,1}^{Q} =0$. Using these to evaluate the static response from Eq. (4), we obtain
\begin{eqnarray}
    & & (v_{0} + v_{1} x + v_{2} x^{2} ) = 1 - \frac{x^{2}}{2} \nonumber\\
\Rightarrow  & & v_{0} = 1; \; v_{1} = 0; \; v_{2} = -\frac{1}{2}.
\end{eqnarray}
Substituting the values of $v_{b}$'s in the expression for pump amplitudes gives 
\begin{subequations}
\begin{align}
  p_{1,0}^{I} &=0; \quad p_{1,1}^{I} = 1\\
    p_{1,0}^{Q} &=0; \quad p_{1,1}^{Q} = 0
\end{align}
    \label{Eqpumpexample1}%
\end{subequations}
and hence the amplitude of first Josephson harmonic as $x \cos\omega_{J} t$. It should be noted that the resultant expression for the static voltage across the junction
\begin{eqnarray}
    \frac{V}{I_{B}R} = \frac{\omega_{J}}{\omega_{B}} = 1 - \frac{x^{2}}{2}
\end{eqnarray}
yields the leading order term from the expansion of the exact expression in the parameter $I_{0}/I_{B}$,
\begin{eqnarray}
    \frac{V}{I_{B}R} = \left(1 - \frac{I_{0}^{2}}{I_{B}^{2}} \right)^{1/2} = 1 - \frac{1}{2} \frac{I_{0}^{2}}{I_{B}^{2}} +\mathcal{O} \left(\frac{I_{0}^{4}}{I_{B}^{4}}\right),
\end{eqnarray}
thus demonstrating the consistency of our evaluation scheme.
Similarly, on going to the next order of expansion (cubic) in
$\delta\varphi$ and including the second Josephson harmonic in the
analysis, we obtain,
\begin{subequations}
\begin{align}
    p_{1,0}^{I} &=0; &  p_{1,1}^{I} &= 1; &  p_{1,2}^{I} &= 0; &  p_{1,3}^{I} &= \frac{1}{4}\\
    p_{1,0}^{Q} &=0;  & p_{1,1}^{Q} &= 0;  & p_{1,2}^{Q} &= 0;  & p_{1,3}^{Q} &= 0 \\
    p_{2,0}^{I} &=0; & p_{2,1}^{I} &= 0; & p_{2,2}^{I} &= 0;  & p_{2,3}^{I} &= 0 \\
    p_{2,0}^{Q} &=0; &  p_{2,1}^{Q} &=0; &  p_{2,2}^{Q} &= -\frac{1}{4}; & p_{2,3}^{Q} &=0.
\end{align}
    \label{Eqpumpexample2}%
\end{subequations}
These terms give an expression for the second Josephson harmonic as $
-\frac{x^{2}}{4} \sin 2\omega_{J} t$. The corresponding expression
for the static voltage is
\begin{eqnarray}
    \frac{V}{I_{B}R} = \frac{\omega_{J}}{\omega_{B}} = 1 - \frac{x^{2}}{2}  - \frac{x^{4}}{8},
\end{eqnarray}
now correct to fourth order of the exact value.
%cccccccccccccccccccccccccccccccccccccccccccccccccccccccc
\section{Scattering coefficients of the RSJ}
%cccccccccccccccccccccccccccccccccccccccccccccccccccccccc
%
To derive the scattering matrix of the RSJ, we follow the method described in \cite{ArchanaSQUID} and first obtain an admittance matrix $\mathbb{Y}_{J}$ of the junction using the pump amplitudes obtained in Eq. (8):
\begin{widetext}
\begin{eqnarray}
    \mathbb{Y}_{J} =
 \left(
    \begin{array}{ccccc}
    0 && - \frac{i x}{2 (1+\Omega_{m})} \left[1 + \frac{x^{2}}{4}\left( \frac{1- \Omega_{m}}{1+\Omega_{m}}\right)\right] &&
    \frac{i x}{2 (1-\Omega_{m})}  \left[1 + \frac{x^{2}}{4}\left(\frac{1+ \Omega_{m}}{1-\Omega_{m}}\right)\right] \\
    \\
    - \frac{i x}{2\Omega_{m}}\left(1 - \frac{x^{2}}{4}\right) && 0 && - \frac{x^{2}}{4 (1- \Omega_{m})}\\
    - \frac{i x}{2\Omega_{m}}\left(1 - \frac{x^{2}}{4}\right) & & - \frac{x^{2}}{4 (1+ \Omega_{m})} && 0
    \end{array}
\right).
\label{EqYmat}
\end{eqnarray}
\end{widetext}
Here, all admittances are normalized with respect to the characteristic admittance $Z_{C}^{-1} = R^{-1}$ corresponding to the shunt conductance. From this, it is straightforward to obtain the scattering matrix of the RSJ, correct to cubic order, by using the identity \cite{Pozar}
\begin{eqnarray}
    \mathbb{S} &=& \mathbb{W}^{-1}. (1+\mathbb{Y}_{J})^{-1}(1-\mathbb{Y}_{J}) .\mathbb{W}
\end{eqnarray}
with  coefficients
\begin{subequations}
\begin{align}
    r_{m} &= 1 + \frac{x^{2}}{1- \Omega_{m}^{2}},\\
    r_{\pm} & = 1 \mp  \frac{x^{2}}{2 \Omega_{m} (1\pm \Omega_{m})},\\
    t_{d} &=
\frac{-i x}{\sqrt{\Omega_{m}(1 + \Omega_{m})}}\left(1 + \frac{x^{2}}{4} \frac{3 + \Omega_{m}^{2}}{1-\Omega_{m}^{2}} \right), \\
    t_{u} & = \frac{-i x}{\sqrt{\Omega_{m}(1 + \Omega_{m})}}\left(1 + \frac{x^{2}}{4} \frac{1 - \Omega_{m}}{1+\Omega_{m}} \right), \\
    s_{d} &= \frac{i x}{\sqrt{\Omega_{m}(1 - \Omega_{m})}}\left(1 + \frac{x^{2}}{4} \frac{3 + \Omega_{m}^{2}}{1-\Omega_{m}^{2}} \right),\\
    s_{u} &= \frac{-i x}{\sqrt{\Omega_{m}(1 - \Omega_{m})}}\left(1 + \frac{x^{2}}{4} \frac{1 + \Omega_{m}}{1-\Omega_{m}} \right),\\
v_{\pm\mp}&= \pm \frac{x^{2}}{2\Omega_{m}} \sqrt{\frac{1 \mp \Omega_{m}}{1 \pm \Omega_{m}}}.
\end{align}
\label{EqRSJScoeffs}%
\end{subequations}
Here, $\Omega_{m} = \omega_{m}/\omega_{B}$, where $\omega_{B} = 2 \pi I_{B} R/\Phi_{0}$, and the diagonal matrix
\begin{eqnarray*}
 \mathbb{W} = {\rm diag}(\sqrt{|\Omega_{m}|}, \sqrt{|\Omega_{m}+\Omega_{J}|}, \sqrt{|\Omega_{m} - \Omega_{J}|})
\end{eqnarray*}
transforms the basis vectors from dimensionless current amplitudes $I[\omega]$ into the relevant photon fluxes $a[\omega] = I[\omega]/\sqrt{\hbar|\omega| R}$. Figure 4(b) is a plot of the relative asymmetry obtained as a difference between net up- and downconverted amplitudes, normalized to the total converted power, calculated from the coefficients in Eqs. (\ref{EqRSJScoeffs}).
\par
The importance of higher pump harmonics is underscored by the degree of the terms that break the symmetry between the up and down frequency conversion amplitudes. The scattering coefficients are symmetric to leading order in the expansion parameter $x$; the lowest order asymmetric term is of order $x^{3}$ as shown by Eqs. (\ref{EqRSJScoeffs}); this necessarily involves higher order mixing products mediated by the second Josephson harmonic of strength $\sim x^{2}$ [Eq. (\ref{Eqpumpexample2})].
%
%cccccccccccccccccccccccccccccccccccccccccccccccccccccccc
\subsection{Fluctuation Spectrum of the RSJ}
%cccccccccccccccccccccccccccccccccccccccccccccccccccccccc
%
%
We now show a representative leading order calculation of the noise spectrum  of the RSJ, using the matrix method introduced here. Consider the admittance matrix derived for the junction in Eq. (\ref{EqYmat}) to leading order in x. The total admittance ($\mathbb{Y}_{T}$) may be calculated as
\begin{eqnarray}
    \mathbb{Y}_{T} = (1 + \mathbb{Y}_{J}),
\end{eqnarray}
where $\mathbb{Y}_{J}$ is normalized with respect to the characteristic admittance $Z_{C}^{-1} = R^{-1}$. We can thus obtain the total impedance matrix of the RSJ, $Z_{T}$ as
\begin{eqnarray}
    Z_{T} = \frac{1}{{\rm det} [\mathbb{Y}_{T}]}
    \left(
    \begin{array}{ccccc}
    1 & & \frac{i x}{2 (1 + \Omega_{m})} & &  \frac{-i x}{2 (1-\Omega_{m})} \\
     \frac{i x}{2  \Omega_{m}} & & 1 & & 0\\
     \frac{i x}{2\Omega_{m}} & & 0 & & 1
    \end{array}
    \right),
    \label{Zmat}
\end{eqnarray}
where we have considered only terms to first order in $x$. The first row of $Z_{T}$ gives the impedance contribution at the modulation frequency $\omega_{m}$. We note that
\begin{eqnarray}
    ({\rm det} [\mathbb{Y}_{T}])^{-1} &=& \left[1 - \frac{x^{2}}{2 (1 - \Omega_{m}^{2})}\right]^{-1} \nonumber\\
    &=& 1 + \frac{x^{2}}{2} + \mathcal{O}(x^{4}).
    \label{EqdynRes}
\end{eqnarray}
Equation (\ref{EqdynRes}) corresponds to the dynamic resistance $R_{D}$ across the junction since, to first order in perturbation, we have
\begin{eqnarray*}
    v_{0} = \frac{V_{\rm dc}}{I_{0}R} =\frac{1}{x} - \frac{x}{2},
    \label{dc}
\end{eqnarray*}
leading to
\begin{eqnarray}
    R^{-1} R_{D} \equiv R^{-1} \frac{d V_{\rm dc}}{d I_{B}} &=& \frac{d v_{0}}{d (1/x)} = 1 + \frac{x^{2}}{2}.
    \label{dynresis}
\end{eqnarray}
Thus, we can write the power spectrum of the voltage fluctuations for the RSJ, in the limit of zero modulation frequency ($\Omega_{m} \rightarrow 0$), as
\begin{eqnarray}
 S_{VV} [\omega_{m}] &=& |z_{11}|^{2} S_{II}[\omega_{m}] + |z_{12}|^{2}S_{II}[\omega_{J}+ \omega_{m}] \nonumber\\
 & & \quad + |z_{13}|^{2}S_{II}[-\omega_{J}+\omega_{m}]\\
 &=& \left[S_{II}[\omega_{m}] + \frac{x^{2}}{4} (S_{II}[\omega_{J}] + S_{II}[-\omega_{J}])\right]  R_{D}^{2} \nonumber\\
 &=& \left[\frac{2 k_{B} T}{R} + \frac{I_{0}^{2}}{I_{B}^{2}} \frac{\hbar \omega_{J}}{2 R} \coth\left(\frac{\hbar \omega_{J}}{2 k_{B} T}\right)\right] R_{D}^{2}. \nonumber\\
\label{noise}
 \end{eqnarray}
Here, in the second step we have used the coefficients in the second row of $Z_{T}$ calculated in Eq. (\ref{Zmat}) and the identity $S_{II}[\omega]+ S_{II}[-\omega] = 2\bar{S}_{II} = 2 R^{-1} \hbar \omega \coth(\hbar \omega/2 k_{B}T)$ \cite{ClerkRMP}. Equation (\ref{noise}) is the well known result \cite{Likharev, SQUIDvol1} showing that the noise appearing near zero frequency in the RSJ involves two contributions: (i) a direct contribution from the input Johnson noise of the resistor and (ii) noise that is downconverted from the Josephson harmonics and appears at the input. The second contribution is appropriately weighted by the square of the bias parameter $x^{2}$. This can be easily understood in the scattering formalism presented here: the strength of the internally generated Josepshon frequency, which acts as the pump for mixing down the high frequency noise, is determined solely by the bias condition. The downconverted amplitude scales directly with the pump strength, a result well known in the literature on parametric systems. Consequently, the intensity of the noise fluctuations, which scale as the square of the amplitude, are weighted by $x^{2}$ [cf. Eq. \ref{Eqpumpexample1}]. Extending this analysis to higher orders provides us with an analytical method to quantify the contribution of higher harmonics and their sidebands to the noise near zero frequency.
\\
%
%cccccccccccccccccccccccccccccccccccccccccccccccccccccccccccccc
\section{Numerical simulation of RSJ mixing}
%cccccccccccccccccccccccccccccccccccccccccccccccccccccccccccccc
%
%
%
\begin{figure}
  % Requires \usepackage{graphicx}
  \includegraphics[width=\columnwidth]{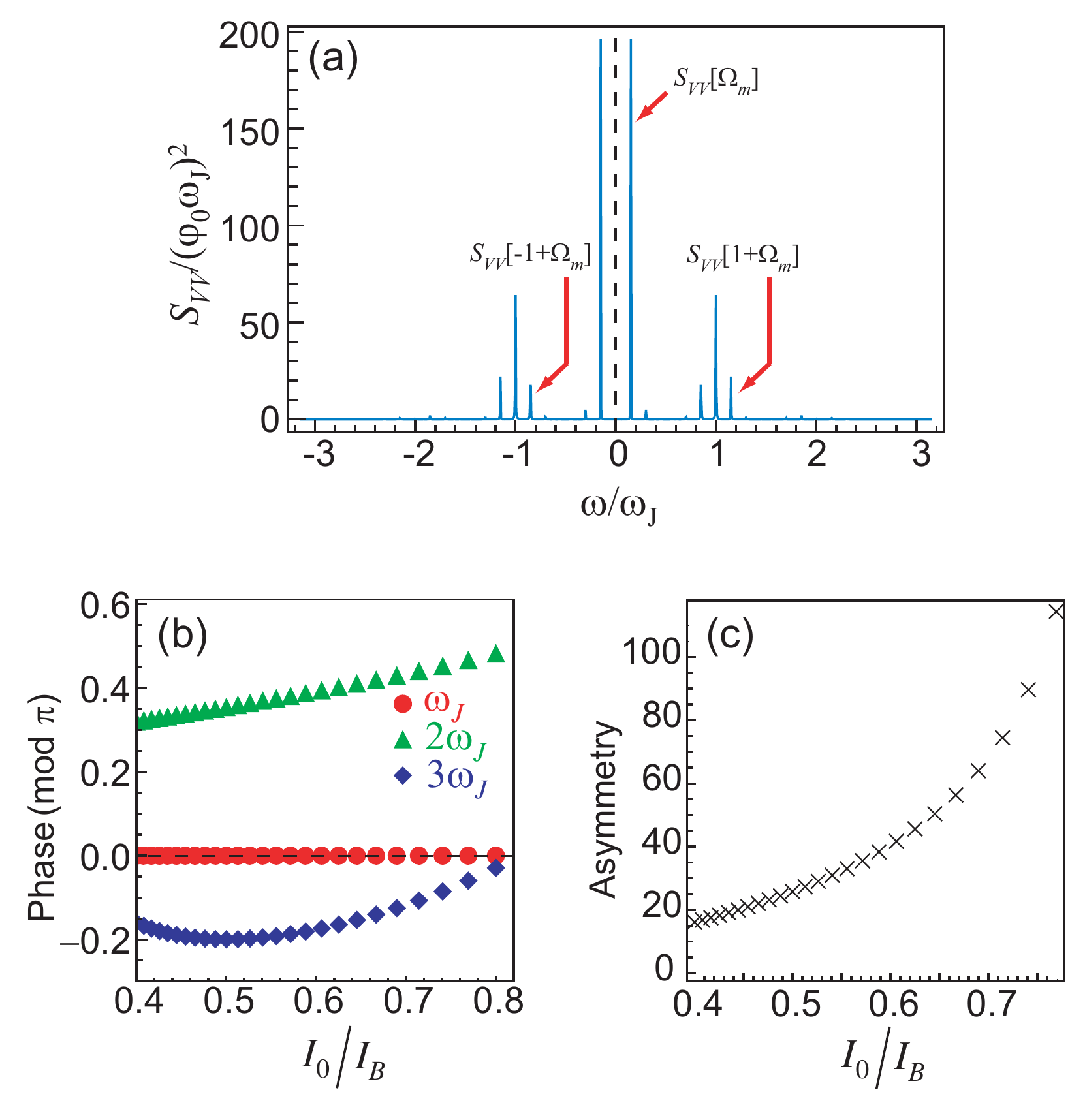}\\
  \caption{Numerical calculation for the RSJ. (a) A representative spectral density of voltage fluctuations across the RSJ, calculated numerically for $I_{B} = 1.5\; I_{0}$. (b) Phases of different Josephson harmonics obtained from a direct numerical integration of Eq. (\ref{EqRSJnum}) with $i_{\rm rf}(\tau)=0$ . (c) The asymmetry obtained from the numerically calculated spectrum, in the presence of an rf source term $i_{\rm rf} (\tau)$ as given in Eq. (\ref{EqSignum}). It is calculated as the difference between the voltage spectral densities at modulation frequency $\Omega_{m}$ and sideband frequencies $(\pm 1 + \Omega_{m})$, respectively, also indicated with red arrows in (a).}
\label{FigRSJnumerics}%
\end{figure}%
One can rewrite Eq. (4) as
\begin{eqnarray}
    \sqrt{i_{B}^{2} -1} \frac{d \varphi}{d\tau} + \sin\varphi = i_{B} + i_{\rm RF}(\tau),
\label{EqRSJnum}
\end{eqnarray}
where $i_{B} = I_{B}/I_{0}$ represents the dimensionless parametrization of the static bias current and $\tau = \omega_{J}t $ represents dimensionless time counted in units of Josephson oscillation period.  The term $i_{RF} (\tau)$ denotes the rf drive at signal and sideband frequencies of interest:
\begin{eqnarray}
 i_{\rm RF}(\tau) &=& m(\tau) \cos(\Omega_{m}\tau) + s_{+}(\tau)\cos [(1+\Omega_{m})\tau] \nonumber\\
& & \qquad \qquad+ s_{-}(\tau) \cos [(1-\Omega_{m})\tau].
\label{EqSignum}
\end{eqnarray}
Here, $\Omega_{m} = \omega_{m}/\omega_{J}$  denotes the dimensionless frequency of the modulation signal and $1 \pm \Omega_{m}$ denote the dimensionless sideband frequencies. To be specific, we assume $m(\tau) = s_{\pm}(\tau) =  0.1$.
\par
Figure \ref{FigRSJnumerics} shows the results obtained by numerical integration of Eq. (\ref{EqRSJnum}), averaged over $\sim 10^{3}$ Josephson periods. An example of a typical spectrum is shown in Fig. \ref{FigRSJnumerics}(a). Figure \ref{FigRSJnumerics}(b) shows that as the bias current $I_{B}$ is decreased towards $I_{0}$, phase configuration of the Josephsonharmonics becomes closer to that given in Eq. (\ref{Eqpumpexample2}). Concomitantly, the asymmetry between the net output powers at the modulation frequency and the sidebands increases with bias current, as shown in Fig. \ref{FigRSJnumerics}(c). The essential physics presented analytically is thus confirmed qualitatively by the numerical calculation. For quantitative comparison, however, one would require a more detailed study that takes into account higher Josephson harmonics ($K > 2$) in the perturbative series expansion, and isolates the effect of reflections and higher order mixing products present in the full numerical calculation.